\documentclass[journal=jacsat,manuscript=article]{achemso}

\usepackage[version=3]{mhchem} 
\usepackage{booktabs}
\usepackage{cleveref}
\usepackage{graphicx}
\usepackage[T1]{fontenc}
\usepackage[utf8]{inputenc}
\usepackage{mathtools}
\usepackage{multirow}
\usepackage{amsmath}
\usepackage{textcomp}
\usepackage{tabularx}
\usepackage[normalem]{ulem}
\usepackage{amssymb}
\usepackage{bm}
\usepackage{xcolor}
\usepackage{float}

\def\etal{{\em et al.\/}}
\def\1Tp{1T$^\prime$}

\author{Marta Loletti}
\affiliation[ICMAB]
{Institut de Ciència de Materials de Barcelona, ICMAB-CSIC, Campus UAB, 08193 Bellaterra, Spain}
\altaffiliation{Contributed equally to this work}

\author{Qi Ren}
\affiliation[ICMAB]
{Institut de Ciència de Materials de Barcelona, ICMAB-CSIC, Campus UAB, 08193 Bellaterra, Spain}
\alsoaffiliation
{School of Aerospace Engineering, Beijing Institute of Technology, Beijing 100081, China}
\altaffiliation{Contributed equally to this work}

\author{Zhuyao Chang}
\affiliation[Hebei Normal University]
{Department of Physics and Hebei Advanced Thin Film Laboratory, Hebei Normal University, Shijiazhuang 050024, China}

\author{Rajan Kumar}
\affiliation[IIT]{Department of Mechanical Engineering, Indian Institute of Technology Bombay, Mumbai, 400076, India}

\author{Nemo McIntosh}
\affiliation[ICMAB]
{Institut de Ciència de Materials de Barcelona, ICMAB-CSIC, Campus UAB, 08193 Bellaterra, Spain}

\author{Mart\'i Raya-Moreno}
\affiliation{Department of Physics and CSMB, Humboldt-Universität zu Berlin, Zum Großen Windkanal 2, 12489 Berlin, Germany}

\author{Ankit Jain}
\affiliation[IIT]{Department of Mechanical Engineering, Indian Institute of Technology Bombay, Mumbai, 400076, India}

\author{Zhao Liu}
\affiliation[Hebei Normal University]
{Department of Physics and Hebei Advanced Thin Film Laboratory, Hebei Normal University, Shijiazhuang 050024, China}

\author{Riccardo Rurali}
\affiliation[ICMAB]
{Institut de Ciència de Materials de Barcelona, ICMAB-CSIC, Campus UAB, 08193 Bellaterra, Spain}
\email{rrurali@icmab.es}

\title{Higher order anharmonicity and polymorphism in the lattice thermal conductivity of transition metal dichalcogenides}

\keywords{transition metal dichalcogenides, phonons, thermal conductivity, four-phonon scattering}

\begin{document}

%

%
%
%

\begin{abstract}
We present a comprehensive {\it ab initio} study of the effect of higher-order anharmonicity and polymorphism on the lattice thermal conductivity, ${\bm \kappa}$, of single-layer transition metal dichalcogenides. We show that four-phonon processes are responsible for an additional suppression of ${\bm \kappa}$ ranging from 15\% to 35\% in the ground-state 2H phase. Conversely, their effect is almost negligible on the thermodynamically competitive \1Tp\ phase. The ${\bm \kappa}$ difference between the 2H higher-conductivity phase and the \1Tp\ lower-conductivity phase can be exploited for the design of thermal switches that may find applications in thermal management and in the development of phonon-based logic elements, particularly in the case of tellurides, where field-controlled ultrafast and reversible phase transitions have already been demonstrated experimentally.
\end{abstract}


\section{Introduction}
\label{sec_intro}

In semiconductors and insulators heat is mostly carried by phonons, the quantized vibration of the crystal lattice. The thermal conductivity, ${\bm \kappa}$, of these materials is determined by the scattering processes that phonons undergo, which limit their propagation. On the one hand, there is scattering with lattice imperfections, such as point defects (including mass defects, i.e. isotopes), stacking faults, and grain boundaries;
on the other hand there are phonon-phonon collisions, resulting from materials' anharmonicity. While the first class of processes can in principle be strongly reduced by controlling the degree of purity of the crystal, the second is a built-in source of scattering that constitutes the true material fingerprint as far as heat transport is concerned.

Four-phonon scattering is usually assumed to play a non-negligible role only at high temperature or in some very specific, highly anharmonic materials. Therefore, perturbative treatments up to the third-order (i.e., considering only collision processes involving three phonons, the lowest degree of anharmonicity) have been generally believed to be accurate enough to predict the lattice thermal conductivity up to room temperature.
However, since methodological advances~\cite{FengPRB15, FengPRB17, HanCPC22, XiaPRB25, LinArXiv25, GuoArXiv25} and the increase in computational power made possible the explicit calculation of higher-order phonon scattering rates, a more nuanced scenario emerged and materials where these effects are already relevant at room temperature were reported~\cite{YangPRB19, RavichandranNatComm19, LiuACSAMI21, MaCPL25, JainJPCL26}.

The case of van der Waals 2D materials is of particular interest in this respect. Feng and Ruan found a significant reduction of the lattice thermal conductivity of single-layer graphene when four-phonon scattering is included, and discuss their results in terms of reflection symmetry selection rules on flexural phonons~\cite{FengPRB18}. Similar conclusions were later reported for h-BN by the same group~\cite{GuoAPL24}, while Tang and coworkers found reductions in ${\bm \kappa}$ of 66 and 62\% at room temperature for TiS$_2$ and TiSe$_2$, respectively~\cite{TangPRB23}.

Transition metal dichalcogenides (TMDs)~\cite{ManzeliNatRevMat17} are one of the most interesting families of 2D materials: they are stable in air, have bandgaps in the visible range that can be tuned by controlling the flake thickness, and find applications in 
electronics~\cite{RadisavljevicNatNano11,WangNatureNano12} optoelectronics~\cite{SplendianiNL10, MakNatPhoto16}, and catalysis~\cite{JaramilloScience07}, among others. Nevertheless, there is no unified picture about the role of higher-order anharmonic phonon processes on the lattice thermal conductivity. Results are either scattered or sometimes even contradictory. For instance, Farris \etal~\cite{FarrisPRB24} found a negligible effect of 4ph scattering in MoS$_2$, with ${\bm \kappa}$ values within 1\% when only three-phonon scattering is considered. On the other hand, Chaudhuri \etal~\cite{ChaudhuriPRB24} reported that introducing 4ph scattering lead to a 79\% decrease of ${\bm \kappa}$. A reduction of the 10\% was found by Gokhale \etal~\cite{GokhalePRB21}, while more recently Kocaba\c{s} \etal~\cite{KocabasAPR25} employed four different machine-learning interatomic potentials (MLIP), leading to reductions of 5, 8, 9, and 16\%, depending on the specific model used. The same four MLIP models lead to reduction of 4, 6, 8, and 20\% for MoSe$_2$~\cite{KocabasAPR25}, while Farris and coworkers~\cite{FarrisPRB24} report an almost negligible 1\% variation (which they also found for WS$_2$, and WSe$_2$). The case of MoTe$_2$ seems to stand out, with a larger 30\% ${\bm \kappa}$ reduction upon inclusion of four-phonon scattering, as reported by Guo \etal\, though relying on the simplified --and usually inaccurate in 2D materials-- relaxation time approximation (RTA)~\cite{GuoMatTodPhys24}.

All the TMDs considered in our study, but WTe$_2$, have in their ground-state a 2H crystal phase, though under certain conditions they can also exhibit the metastable \1Tp\ polymorph. It is a distorted version of the 1T phase, where each Mo atom is coordinated by 6 S atoms in an octahedral geometry, as opposed to the trigonal prismatic coordination in the 2H phase, and can often be energetically competitive with the 2H ground-state~\cite{SokolikovaChemSocRev20}. Illustrative is, in this sense, the case of MoTe$_2$, whose energy barrier between the 2H and \1Tp\ phases is barely 30-40~meV/f.u.~\cite{DuerlooNatComm14, HiddingACSPhoto24, ManojKumarJAP25}, so that both polymorphs have been reported to be almost equally probable depending on the synthesis conditions~\cite{ParkACSNano15, YooAdvMat17}, thereby limiting a tight control on material properties. Reports on the role of 4ph scattering in TMDs are scarce and are limited to group-IVB TMDs of the type MX$_2$ (M= Ti, Zr, Hf; X = S, Se)~\cite{TangPRB23}, with very significant reductions of cubic thermal conductivity that can reach 60-70\%, and ReS$_2$~\cite{YangAPL24}.

In this paper we present an comprehensive study, within an entirely {\it ab initio} theoretical framework, of the lattice thermal conductivity of six single-layer group-VIB TMDs, MX$_2$ (M= Mo, W, X = S, Se, Te) including three-phonon (3ph) and four-phonon (4ph) scattering processes, and considering both the 2H ground-state and \1Tp\ metastable phase. Because of their synthesis maturity (amenable to both CVD growth and mechanical exfoliation), indirect-to-direct bandgap enabling strong photoluminescence, stability in air, natural abundance and low cost, this is by far the most appealing class of TMDs for several applications. Our main focus is assessing quantitatively to what extent 4ph processes are important and how much they reduce the computed thermal conductivity when only accounting for 3ph scattering. The use of the same methodology and computational parameters ensures all materials are treated on equal footing, considerably limiting the possibility that observed differences stem from computational artifacts, rather than true material properties. 

\section{Methods}
\label{sec:methods}

The optimized geometry, as well as the harmonic and anharmonic interatomic force constants (IFCs), were obtained from density-functional calculations with the \texttt{VASP} code~\cite{KressePRB93, Kresse94, KressePRB96, KresseCMS96}, using the projector augmented wave method~\cite{BlochlPRB94, KressePRB99}, and the Local Density Approximation (LDA) for the exchange-correlation energy. We studied the 2H and \1Tp\ polymorphs, first optimizing the atomic positions and the in-plane lattice vectors until forces and stress were lower than $10^{-3}$~eV/\AA\ and $2 \times 10^{-2}$~kbar, respectively; the $c$-vector was kept fixed to 25~\AA, allowing a vacuum buffer of $\approx$~18~\AA\ to separate the single-layer to its periodic images. The Brillouin zone was sampled with a grid of $24 \times 24$ and $14 \times 26$ {\bf k}-points for the 2H and \1Tp\ phase, respectively.

The thermal conductivity is obtained by solving the linearized phonon Boltzmann Transport Equation (BTE) using FourPhonon~\cite{HanCPC22}, an extension of the ShengBTE code~\cite{LiCPC14}, after enforcing the rotational invariance of the crystal symmetry in order to guarantee the quadratic dispersion of the lowest phonon branch~\cite{GazisPRB66, WangJPCM07, CarreteMRL16}. In this framework, both three-phonon (3ph) and four-phonon (4ph) scattering processes are accounted for and treated beyond the Relaxation Time Approximation (RTA). We also compute the coherence contribution to the thermal conductivity, solving the Wigner Transport Equation as implemented in an in-house modified version of the \texttt{almaBTE} code~\cite{mrm24_almabte_4ph} to make sure that it is negligible, as expected in well-ordered single crystals like those studied here. For comparison and as a check of our implementation, we provide the results for La$_2$Zr$_2$O$_7$ (see Figure S11 in the Supporting Information) , which are in excellent agreement with those of Ref.~\citenum{SimoncelliPRX22}.

\section{Results and Discussion}
\label{sec:results}

\subsection{2H ground-state}
\label{sub:2h}

\begin{figure}[t]
\includegraphics[width=1.00\linewidth]{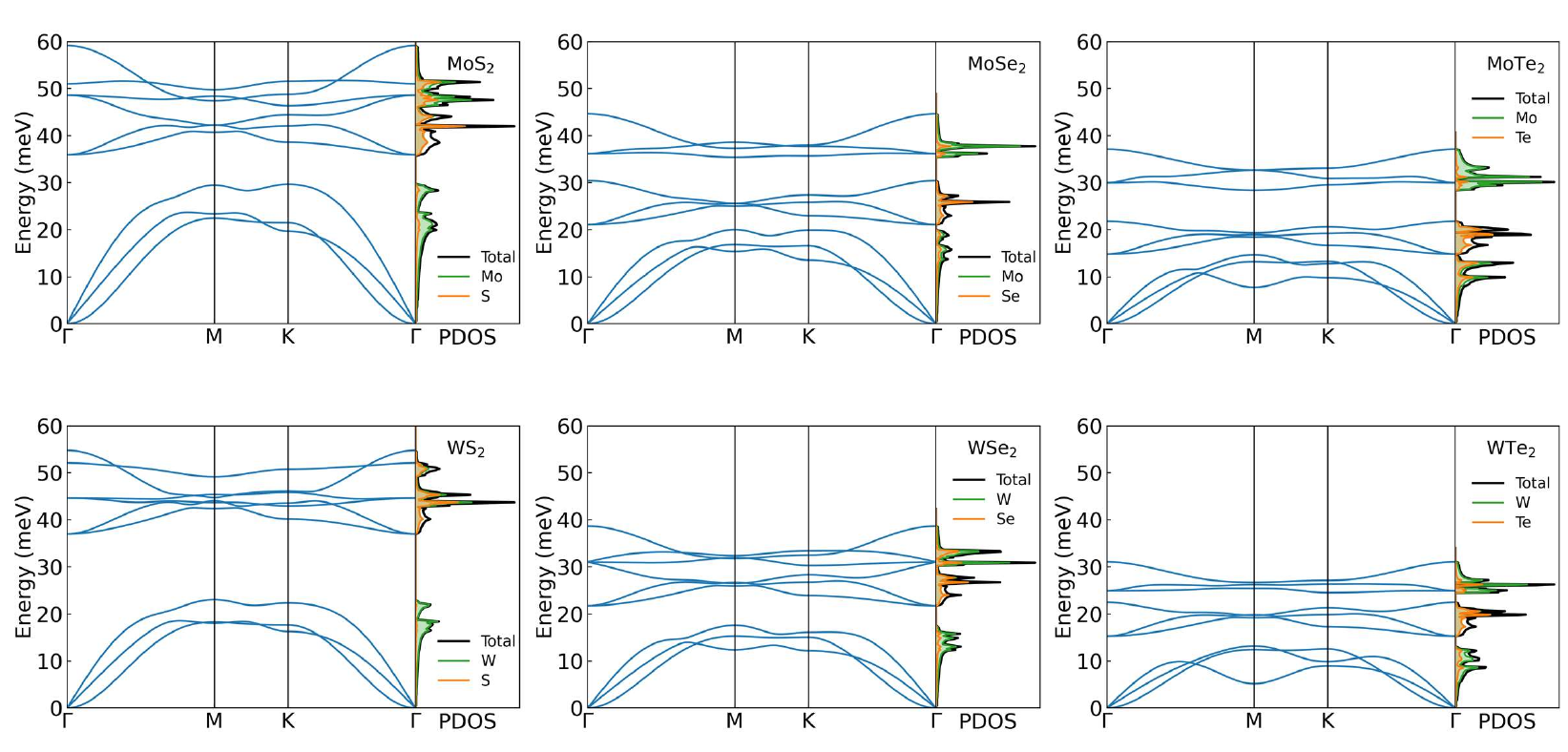}
\caption{Phonon dispersion and density of states of (top row, left to right) 2H MoS$_2$, MoSe$_2$, MoTe$_2$, and (bottom row, left to right) 2H WS$_2$, WSe$_2$, WeTe$_2$.}
\label{fig:bands}
\end{figure}

The phonon dispersions, shown in Figure~\ref{fig:bands}, feature no imaginary mode, thus indicating the expected dynamical stability of all six materials. Sulphides exhibit the larger phonon frequencies, as they involve the lighter chalcogen atom, followed by selenides and tellurides. Metal atoms have a similar effect, though less pronounced, thus a compound with the lighter Mo has larger phonon frequencies than its counterpart with W. Atomic masses are also important to determine the acoustic-optic (AO) bandgap: the closer to unity the mass ratio, the smaller the AO gap. Hence, MoTe$_2$, with a mass ratio $m_\text{Te}/m_\text{Mo} = 1.3$ has virtually no AO gap, while WS$_2$, with a mass ratio $m_\text{S}/m_\text{W} = 0.17$, exhibits the larger AO gap. Interestingly, these considerations that are exclusively based on harmonic properties, already provide a rough interpretive framework of the thermal conductivities: higher phonon frequencies imply higher phonon velocities (see Figure S1 in the Supporting Information), thus higher ${\bm \kappa}$; large AO gap result in a smaller phase-space (i.e., less permitted phonon-phonon processes), also giving a higher ${\bm \kappa}$.
Therefore, sulphides are expected to have the larger ${\bm \kappa}$; MoS$_2$ should be more conductive than WS$_2$ (Mo is lighter than W), but has a smaller AO gap, and thus the latter turns out to have a larger ${\bm \kappa}$. Along the same lines, MoTe$_2$ involves the heavier chalcogen atoms among those considered and has also no AO gap, thus it is expected to be the least conductive.

\begin{figure}[t]
\includegraphics[width=1.00\linewidth]{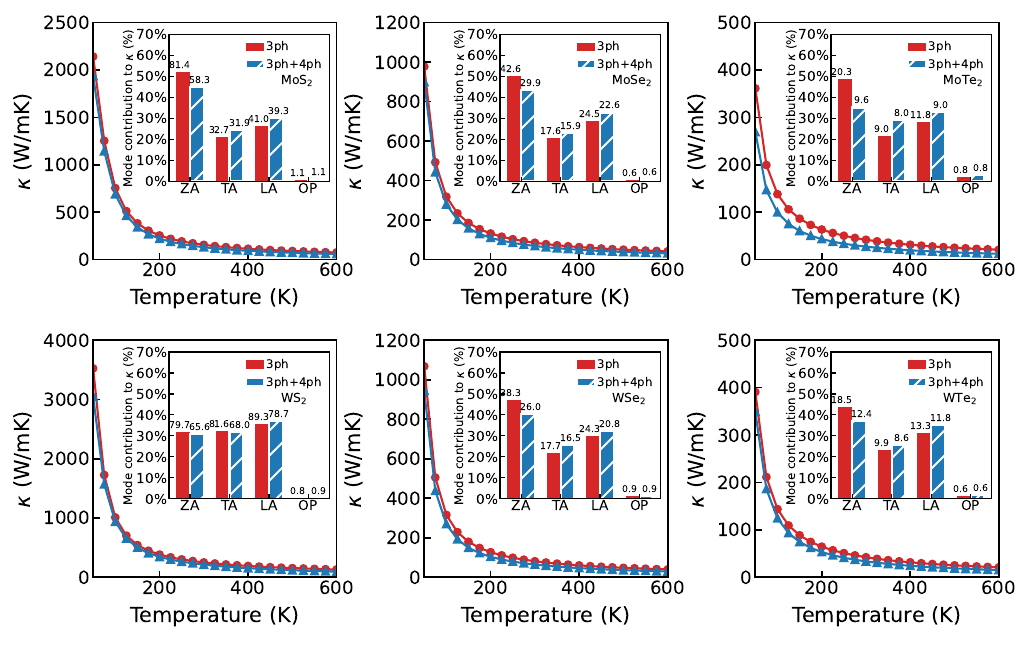}
\caption{Thermal conductivity as a function of temperature of (top row, left to right) 2H MoS$_2$, MoSe$_2$, MoTe$_2$, and (bottom row, left to right) 2H WS$_2$, WSe$_2$, WeTe$_2$ when considering 3ph (red circles) and 3ph+4ph scattering processes (blue triangles). The insets display the mode contribution to ${\bm \kappa}$ at room temperature; the number above the column indicates the absolute value, the height of the column the relative weight of that component over the total ${\bm \kappa}$}
\label{fig:kappa}
\end{figure}

These predictions are qualitatively verified by the computed thermal conductivities, displayed in Figure~\ref{fig:kappa}. As we are primarily interested in the role of 4ph scattering, we plot separately ${\bm \kappa}_{3ph}$, where only third-order anharmonic processes are considered, and ${\bm \kappa}_{3ph+4ph}$, where anharmonic processes up to the fourth-order (i.e., 3ph and 4ph) are accounted for. In the inset of each panel we also present the mode resolved thermal conductivity at 300~K, where we distinguish contributions from flexural acoustic (ZA), transverse acoustic (TA), longitudinal acoustic (LA), and optic phonon modes (OP).
A first notable feature highlighted by Figure~\ref{fig:kappa} concerns WS$_2$, where room temperature ${\bm \kappa}$ has equal contributions from all acoustic modes, while ZA mode is dominant in all the other five 2H-TMDs. A perhaps more important observation, however, is the the stronger reduction of the thermal conductivity in MoTe$_2$ caused by 4ph processes. Such a higher-order anharmonic suppression of ${\bm \kappa}$ has already been reported~\cite{GuoMatTodPhys24}. However,  the comparative analysis of Figure~\ref{fig:kappa} allows concluding that MoTe$_2$ represents an exception in this respect within single-layer group-VIB 2H-TMDs. A closer look at the mode decomposed ${\bm \kappa}$, shown in the inset for of $T=300$~K, reveals that the anomalous behavior of MoTe$_2$ mostly stems from the reduction of the ZA contribution, which drops from 20.3 to 9.6~W~m$^{-1}$K$^{-1}$, a 53\% decrease, followed by LA and TA, whose contribution drops 24 and 11\%, respectively. For comparison, the largest decrease among the other 5 single-layer TMDs investigated is 33\% for ZA modes of WTe$_2$.


\begin{figure}[h]
\includegraphics[width=0.75\linewidth]{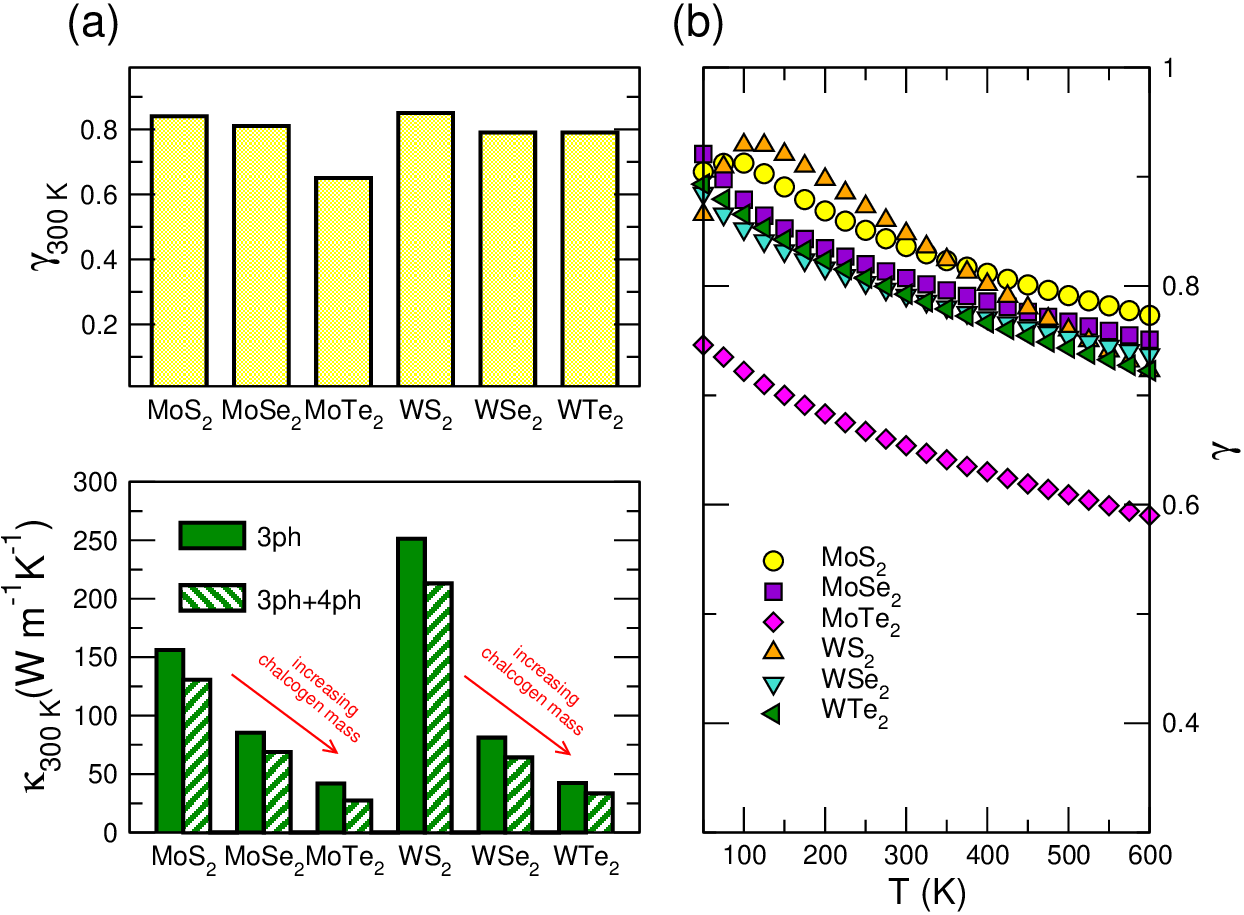}
\caption{(a) Room temperature thermal conductivity considering only 3ph processes, $\kappa_{3ph}$, and both 3ph and 4ph processes, $\kappa_{3ph+4ph}$, and ratio between them, $\gamma = \kappa_{3ph+4ph}/\kappa_{3ph}$ of 2H TMDs. (b) Ratio between $\kappa_{3ph+4ph}$ and $\kappa_{3ph}$ as a function of temperature.}
\label{fig:4phred}
\end{figure}


\begin{figure}[h]
\includegraphics[width=1.00\linewidth]{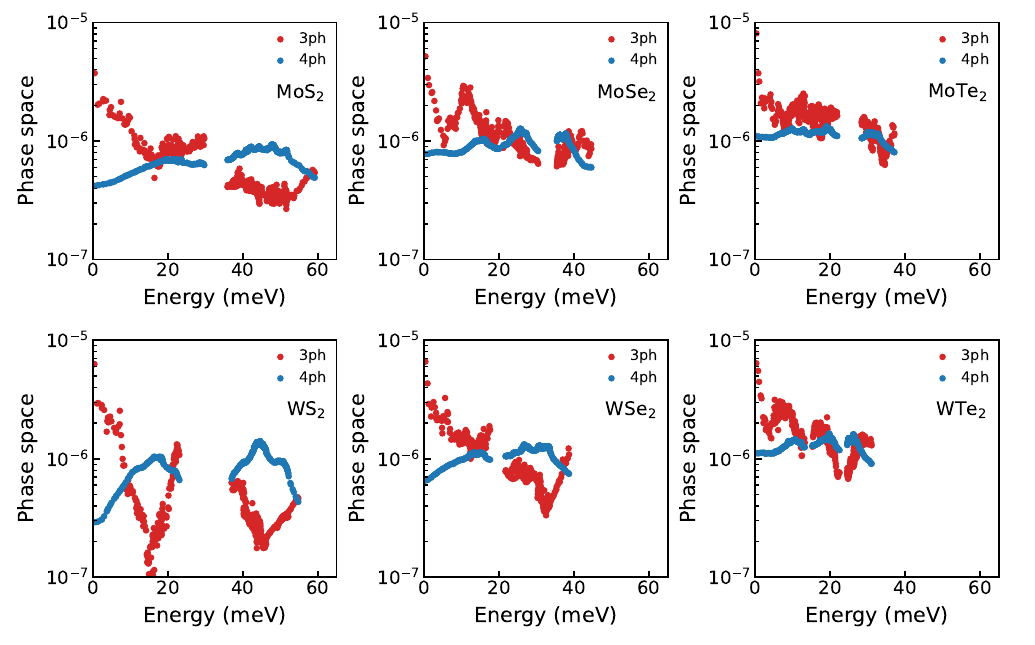}
\caption{Phase-space of 3ph and 4ph processes as a function of phonon energy of (top row, left to right) 2H MoS$_2$, MoSe$_2$, MoTe$_2$, and (bottom row, left to right) 2H WS$_2$, WSe$_2$, WeTe$_2$.}
\label{fig:phasespace}
\end{figure}

\begin{figure}[h]
\includegraphics[width=1.00\linewidth]{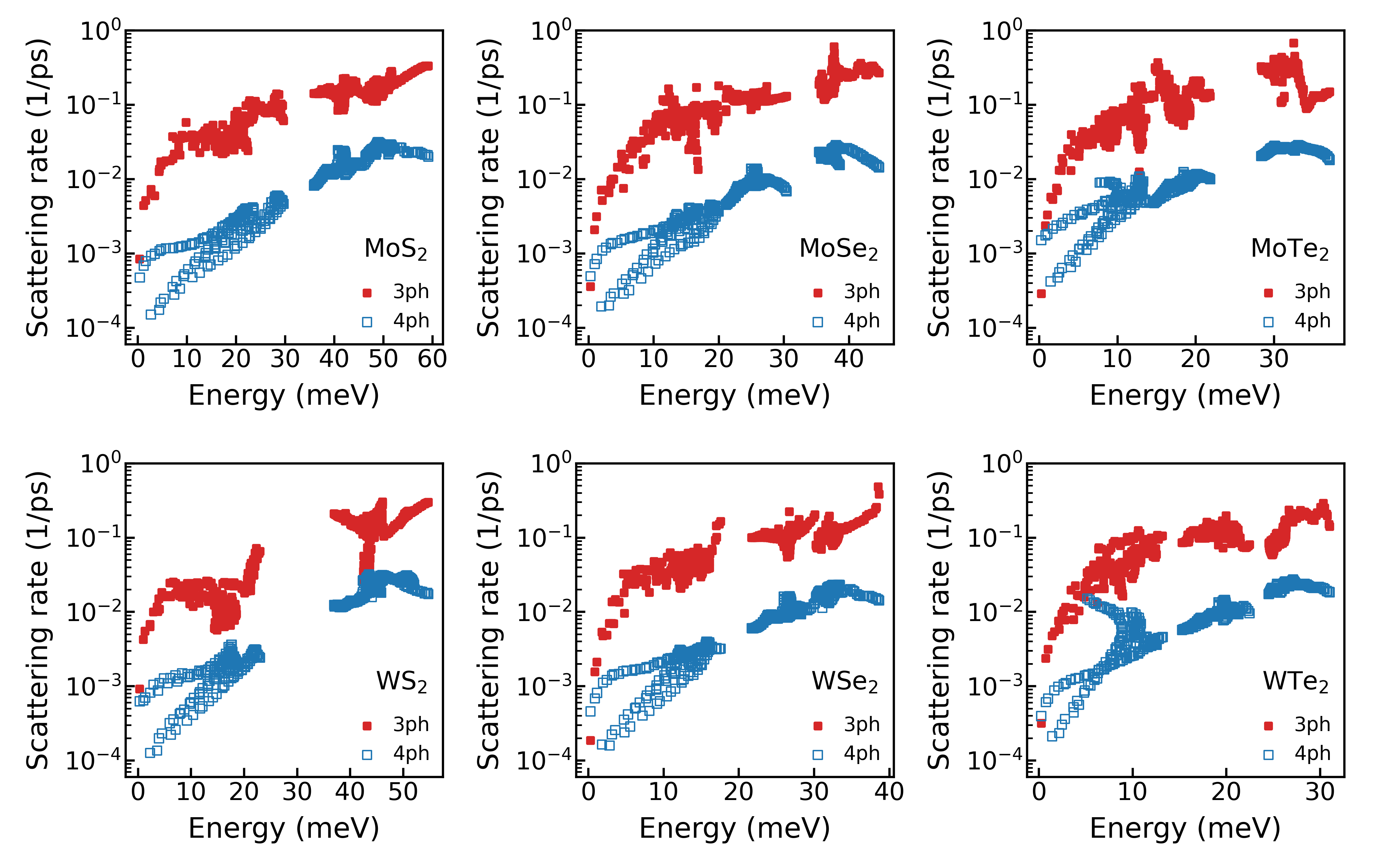}
\caption{Scattering rates of 3ph and 4ph processes as a function of phonon energy of (top row, left to right) 2H MoS$_2$, MoSe$_2$, MoTe$_2$, and (bottom row, left to right) 2H WS$_2$, WSe$_2$, WeTe$_2$.}
\label{fig:scattrates}
\end{figure}

We now take a closer look to the role of 4ph scattering in all materials. In Figure~\ref{fig:4phred}a we show the room temperature thermal conductivity considering only 3ph or 3ph and 4ph processes and the corresponding reduction, $\gamma = \kappa_{3ph+4ph}/\kappa_{3ph}$. As one can see, the decrease of ${\bm \kappa}$ due to 4ph scattering is non-negligible and ranges from 15\% (for WS$_2$) to 35\% (for MoTe$_2$). Four-phonon processes become increasingly more efficient at high-T and indeed we observe a decrease of $\gamma$ as temperature rises, as shown in Figure~\ref{fig:4phred}b. Notice, though, that even at 50~K, the lowest temperature considered, accounting for 4ph process already results in $\approx$8-13\% decrease, which peaks at roughly 25\% for MoTe$_2$.

To shed light on the role of 4ph scattering, we have computed the 3ph and 4ph phase-space and scattering rates. The phase-space, shown in Figure~\ref{fig:phasespace}, is the set of allowed phonon-phonon scattering processes, those conserving energy and momentum and usually exhibits a negative correlation with ${\bm \kappa}$: if the phase-space is large, many scattering channels are available, creating the conditions (but without guaranteeing it) for strong phonon-phonon scattering. In general, we observe an increase of both the three- and the four-phonon phase-space as the ratio between the mass of the chalcogen atom and the metal atom approaches one, as this factor roughly determines the AO gap. Therefore, within the Mo-based TMDs the AO gap decreases and tends to vanish as one moves from S, to Se, and to Te, leading to an increase in the phase-space, as smaller AO gap favors the possibility of phonon-phonon processes that involve both acoustic and optic modes. This trends is perhaps clearer in the weighted phase-space at room temperature, where only the phonon effectively populated at a given temperature are considered (see Figure S2 in the Supporting Information). We find a similar trend for the W-based compounds. However, the phase-space does not seem to be the reason behind the stronger suppression of ${\bm \kappa}$ due to 4ph processes observed in MoTe$_2$. While it can be argued that the available phase-space for 4ph process is larger for tellurides than for sulphides and selenides, there is no substantial differences between MoTe$_2$ and WTe$_2$ (actually, the 4ph phase-space of WTe$_2$ is slightly larger; see Figure S2 in the Supporting Information for a direct comparison between the 4ph phase-space and the 4ph weighted phase-space of the two tellurides).

We now move to the analysis of 3ph and 4ph scattering rates, displayed in Figure~\ref{fig:scattrates}. As one can see there, 3ph scattering processes are roughly one order of magnitude more intense than 4ph processes. There are, however, important exceptions to this general trend. If we focus on MoTe$_2$, which has the largest effects due to 4ph collisions, we see a significant fraction of low-frequency 4ph processes whose scattering rates largely exceeds 10$^{-3}$~ps$^{-1}$, becoming comparable to 3ph scattering rates. Something similar happens with WTe$_2$ and WSe$_2$. However, (i)~in WTe$_2$ 4ph scattering rates become comparable to 3ph ones at higher phonon energies, $\approx 4$~meV, where phonon velocities are lower and thus the overall contribution to ${\bm \kappa}$ is also lower; (ii)~in WSe$_2$ 4ph scattering rates approach 3ph ones in the very low energies limit, but to a lesser extent compared to what observed in MoTe$_2$.

In the Supporting Information we also show plots of the phonon velocity (Figure S1), the weighted phase-space at room temperature (Figure S2), the Gr\"uneisen parameter (Figure S3) and the cumulative thermal conductivity (Figure S4), all as a function of the phonon energy.

\subsection{\1Tp\ metastable phase}
\label{sub:1tp}

Despite being generally metastable with respect to the 2H phase, the \1Tp\ polymorph has been reported experimentally in all six TMDs studied in this work~\cite{LiuNatMat18, YuNatChem18, SongSciAdv23, ChenAdvMat21, LiuNE18, ParkACSNano15, SokolikovaNatComm19, YaoAPL25, Naylor2DM17} and is thus important to characterize their thermal transport properties. The stability of the \1Tp\ phase increases when the chalcogen moves from S to Se to Te. Therefore, while sulfides strongly favor the 2H phase and the \1Tp\ is highly metastable, in selenides the \1Tp\ phase gets closer in energy to the 2H, though still metastable. On the other hand, in tellurides both polymorphs are essentially equally stable, and the \1Tp\ is the thermodynamical stable phase for WTe$_2$, while in MoTe$_2$ both phases can be obtained with the same probability, depending on the synthesis conditions~\cite{ParkACSNano15, YooAdvMat17}. Indeed, we obtain $\Delta E = E_\text{2H}- E_\text{1T$^\prime$}= -0.61$, $-0.37$, and $-0.04$~eV/f.u. for MoS$_2$, MoSe$_2$, MoTe$_2$, and $-0.60$, $-0.32$, and $0.07$~eV/f.u. for WS$_2$, WSe$_2$, WTe$_2$.
Nevertheless, even in those cases where the 2H is clearly energetically favored, the \1Tp\ phase can be stabilized by chemical intercalation~\cite{TanJACS17, WangAdvEnSustRes21, NongJACS25}, electrostatic doping~\cite{ZhuangPRB17, WangAdvFunctMat18}, strain~\cite{WangAdvFunctMat18, ZanASS22, AwateACSNano23, TaoJACS24}, or electron/ion irradiation~\cite{ChoScience15, ZhuJACS17}.

\begin{figure}[t]
\includegraphics[width=1.00\linewidth]{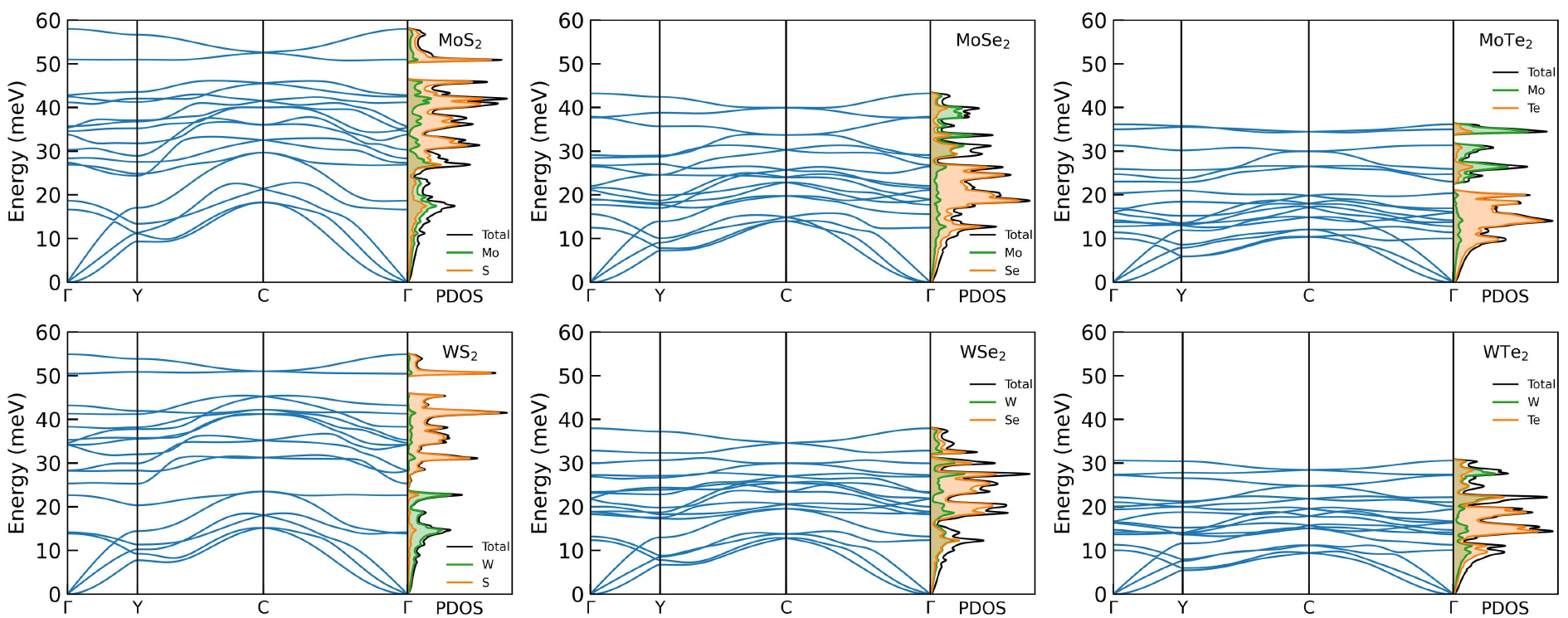}
\caption{Phonon dispersion and density of states of (top row, left to right) \1Tp\ MoS$_2$, MoSe$_2$, MoTe$_2$, and (bottom row, left to right) \1Tp\ WS$_2$, WSe$_2$, WeTe$_2$.}
\label{fig:bands_1Tp}
\end{figure}

The phonon dispersions of the six \1Tp\ single-layer TMDs is displayed in Figure~\ref{fig:bands_1Tp} and also for this polymorph we do not find any imaginary phonon bands, indicating that all these materials are dynamically stable. One thing that immediately stands out is that, by lifting several degeneracies, the reduced symmetry of the \1Tp\ phase results into many more phonon bands that occupy almost uniformly the bandwidth, with no clearly defined AO bandgaps, as observed in the 2H phase, particularly in sulphides and selenides (see Figure~\ref{fig:bands}). By itself, this fact should lead to a reduced thermal conductivity, driven by an enlarged phase-space: the absence of an AO bandgap increases the probability that optical phonon scatter acoustic phonons, which usually carry most of the total heat.

\begin{figure}[t]
\includegraphics[width=1.00\linewidth]{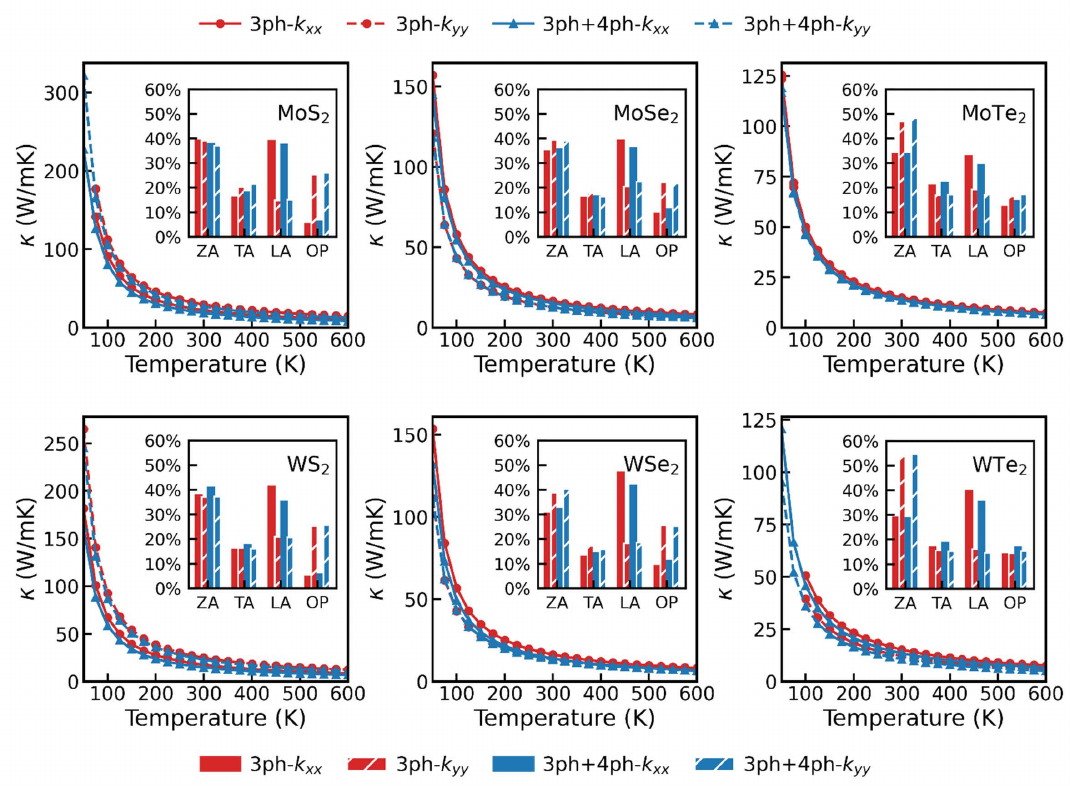}
\caption{Thermal conductivity as a function of temperature of (top row, left to right) \1Tp\ MoS$_2$, MoSe$_2$, MoTe$_2$, and (bottom row, left to right) \1Tp\ WS$_2$, WSe$_2$, WeTe$_2$ when considering 3ph (red circles) and 3ph+4ph scattering processes (blue triangles). The insets display the mode contribution to ${\bm \kappa}$ at room temperature; the number above the column indicates the absolute value, the height of the column the relative weight of that component over the total ${\bm \kappa}$}
\label{fig:kappa_1Tp}
\end{figure}

These qualitative considerations are confirmed by the computed thermal conductivities, displayed in Figure~\ref{fig:kappa_1Tp}, where, like in the case of the 2H phase, we plot separately ${\bm \kappa}_{3ph}$ and ${\bm \kappa}_{3ph+4ph}$ and where in the inset we show the mode resolved thermal conductivity at 300~K, separating the contributions from ZA, TA, LA, and OP modes. As one can see there, for all six materials we obtain thermal conductivities that are considerably lower than the 2H counterpart. Notice also that, due to the reduced symmetry of the \1Tp\ phase, the in-plane transport directions are no longer equivalent and $\kappa_{xx} \neq \kappa_{yy}$, thus each panel features two curves for ${\bm \kappa}_{3ph}$ and two curves for ${\bm \kappa}_{3ph+4ph}$. As inferred by the visual inspection of the phonon dispersions, the reduced thermal conductivity mostly stems from the increased suppression of acoustic phonons, as can be appreciated in the figure's insets, where the relative fraction of heat contributed by OP is now sizeable, often of the same order of that of TA. In the 2H phase, on the other hand, acoustic phonons were much more efficient heat carriers and the relative contribution of OP modes was usually negligible (see the insets of Figure~\ref{fig:kappa}).

\begin{figure}[h]
\includegraphics[width=0.75\linewidth]{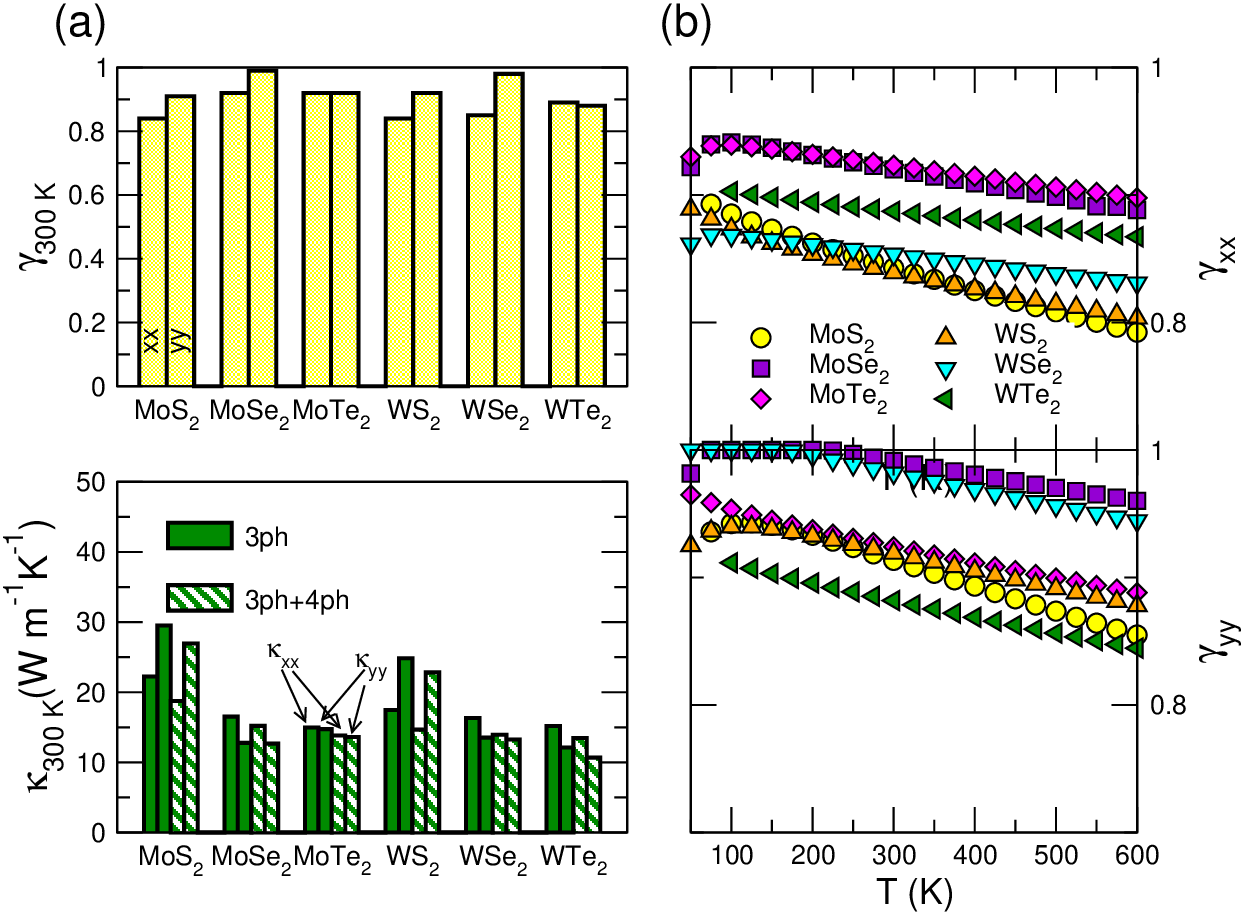}
\caption{(a) Room temperature thermal conductivity considering only 3ph processes, $\kappa_{3ph}$, and both 3ph and 4ph processes, $\kappa_{3ph+4ph}$, and ratio between them, $\gamma = \kappa_{3ph+4ph}/\kappa_{3ph}$ of \1Tp\ TMDs. (b) Ratio between $\kappa_{3ph+4ph}$ and $\kappa_{3ph}$ as a function of temperature, for transport along the )top) $x$- and (bottom) $y$-axis.}
\label{fig:4phred_1Tp}
\end{figure}

Another important difference from the 2H phase is that in the \1Tp\  polymorph phonon dynamics is largely determined by third-order anharmonic processes and 4ph scattering have a smaller effect on the total thermal conductivity (see Figure~\ref{fig:4phred_1Tp}). In sulphides and in WSe$_2$ the room temperature value of $\gamma_{xx}$ reaches 16\%, but in other cases the 4ph reduction is less than 10\% (0.99 for transport along the $y$-axis in MoSe$_2$). It is also noteworthy that 4ph collisions have a different effect, within the same material, when looking at $\kappa_{xx}$ or $\kappa_{yy}$, thus altering the degree of anisotropy, $\kappa_{xx}/\kappa_{yy}$. For instance, 4ph processes bring upon a reduction, as compared to the case when only 3ph processes are considered, of 15\% to $\kappa_{xx}$ in WSe$_2$, while  $\kappa_{yy}$ is almost unchanged (2\% reduction). Similar conclusions are drawn when considering the temperature dependence of $\gamma$, so while MoSe$_2$ has one of the $\kappa_{yy}$ that is less affected by 4ph scattering ($\gamma_{yy} \approx 0.96-1$), with almost no reduction from 4ph scattering up to $\approx 200$~K, its $\kappa_{xx}$ can be reduced as much as $\approx 10$\% ($\gamma_{xx} \approx 0.89-0.92$). The reason for the observed behavior lies in the increased phase-space for three-phonon processes, driven by the reduced symmetry: third-order anharmonic processes become so efficient that higher-order processes have little effect.

As for anisotropy, it is worth noting that it is material dependent. Sulphides features the largest differences between $\kappa_{xx}$ and $\kappa_{yy}$, and the more conductive direction is the $y$-axis. Conversely, in selenides and tellurides  $\kappa_{xx} > \kappa_{yy}$, but with smaller differences (MoTe$_2$ is essentially isotropic).

In the Supporting Information we also show plots of the phonon velocity (Figure S5), phase-space (Figure S6), the weighted phase-space at room temperature (Figure S7), the scattering rates (Figure S8), the Gr\"uneisen parameter (Figure S9) and the cumulative thermal conductivity (Figure S10), all as a function of the phonon energy.

\subsection{Thermal switching via phase transition}
\label{sub:switch}

Polymorphism is an important asset that TMDs bring to the table, because by controlling the crystal phase one can design materials with, to certain extent, tailor-made properties. As discussed above, the 2H phase is the thermodynamical ground-state for all the materials studied here, but WTe$_2$, who slightly favors the \1Tp\ phase. However, all materials have been reported in both crystal phases and the metastable phase can be stabilized via e.g. strain~\cite{AwateACSNano23} or chemical intercalation~\cite{WangAdvEnSustRes21}. Of particular interest are those TMDs where the energy barrier separating the two phases is moderate and that are thus amenable to field-controlled phase transitions, paving the way to the dynamical control of material's properties. This is the case of MoTe$_2$, whose energy barrier for the transition between the 2H and \1Tp\ phases~\cite{SokolikovaChemSocRev20} has been experimentally estimated to be 30-40~meV/f.u.~\cite{DuerlooNatComm14, HiddingACSPhoto24, ManojKumarJAP25}, much smaller e.g. than 800~meV/f.u. for MoS$_2$ and 250~meV/f.u. for MoSe$_2$ (see e.g. Ref.~\cite{DuerlooNatComm14}).
For these reason, polymorphism is MoTe$_2$ --but also in other TMDs, provided field-induced phase transitions exist-- can be exploited to achieve a dynamical modulation of the thermal conductivity --a feature that would prove critical for many applications, ranging from phonon-based logic~\cite{LiRMP12, NatafNatRevMat24} to energy harvesting~\cite{WehmeyerAPR17}. Therefore, we take a closer look to the transition between the 2H and \1Tp\ crystal phases and discuss the possibility to exploit it to design thermal switches, where the access to a low- and high-conductivity thermal conduction state can be controlled with an external stimulus.

\begin{figure}[h]
\includegraphics[width=0.75\linewidth]{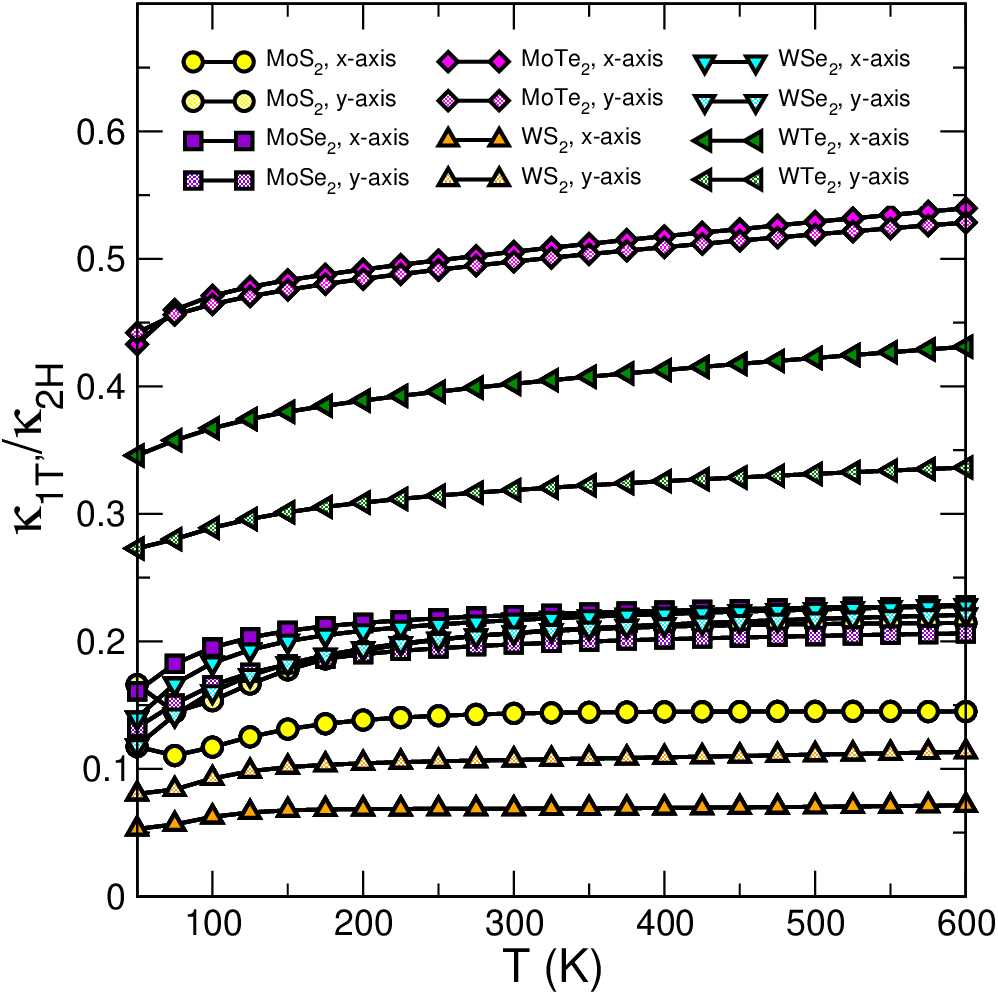}
\caption{Ratio between the thermal conductivity of the two crystal phases, $\kappa_{1T^\prime}/\kappa_{2H}$ as a function of temperature.}
\label{fig:switch}
\end{figure}

Our results are summarized in Figure~\ref{fig:switch}. As one can see, sulphides and selenides features an extraordinary reduction of the thermal conductivity upon the 2H$\rightarrow$\1Tp\ phase transition, with a ${\bm \kappa}$ suppression exceeding 90\% for WS$_2$. Unfortunately, however, these are also the materials where the barriers that must be overcome are higher. Therefore, although both polymorphs can be synthesized/stabilized, realizing reversible, field-induced phase transitions is less straightforward. Tellurides represent a more interesting case for applications. The reduction $\kappa_{1T^\prime}/\kappa_{2H}$ is considerable, ranging from 49\% (MoTe$_2$) to 68\% (WTe$_2$) at room-temperature, but the energy barriers are lower. Indeed, the dynamical and reversible phase transition between the 2H and the \1Tp\ polymorph in MoTe$_2$ has been demonstrated by means of different external stimuli, including electric field~\cite{WangNature17, HeMH22}, mechanical strain~\cite{SongNL16}, laser irradiation~\cite{TanNanoscale18}, Joule heating~\cite{ManojKumarJAP25}, and light absorption~\cite{SiNL19, Pengnpj2D20}.

\subsection{Coherent contribution to thermal transport}
\label{sub:coh}

The results presented up to this point are based on the solution of the linearized BTE, which treats phonons as particles that scatter off each other (and off defects and boundaries, when present). This approach, usually accurate for single-crystalline materials, neglects coherent interactions between phonons and only considers particle-like collisions. Disordered materials --e.g. amorphous solids, glasses-- or materials with a complex crystal structure may feature phonon modes that are closely spaced in energy and can thus interfere with one another through a wave-like tunneling process. This behavior, unlike conventional BTE, is captured by Wigner transport theory~\cite{SimoncelliPRX22}.

Although in simple, well-ordered crystals this coherence term is generally negligible and thus is not expected to play an important role in single-layer TMDs, we have verified that this is indeed the case and quantified the difference, if any, between the 2H and the \1Tp\ phase. To this end, we have implemented Wigner theory at the Wigner-RTA level (i.e., assuming a complete decoupling between coherence and occupations, while the occupation part can then be treated either within the BTE RTA or the linearized BTE level) in almaBTE~\cite{CarreteCPC17}, accounting for isotopic, 3ph and 4ph scattering processes.
Our results at room temperature are collected in Table~\ref{tab:wigner}.
As expected for such simple crystals, the coherence component at room temperature is several orders of magnitude smaller than the Peierls component. Consequently, intraband contributions --i.e., coherence effects-- have a negligible impact on the total thermal conductivity. As expected, although still negligible, is somewhat larger in more complex \1Tp\ phase, which has a few nearly-degenerate phonon branches due to its lower symmetry.

\vskip 10pt
 \begin{table}[h]
    \centering
    \renewcommand{\arraystretch}{1.3} 
    \begin{tabular}{l c c c c c c }
        \hline \hline
                    & 2H \\
        \hline
                    & ~~MoS$_2$~~ & ~~MoSe$_2$~~ & ~~MoTe$_2$~~ & ~~WS$_2$~~ & ~~WSe$_2$~~ & ~~WTe$_2$~~ \\
        \hline
        $\kappa_{xx}^{P}$(300~K)~~ & 102.00 & 57.58 & 24.73 & 121.99 & 51.13 & 30.36 \\
        $\kappa_{xx}^{coh}$(300~K)~~ & 0.01 & 0.02 & 0.01 & 0.01 & 0.01 & 0.01 \\
        \hline\hline
                    & \1Tp \\
        \hline
                    & ~~MoS$_2$~~ & ~~MoSe$_2$~~ & ~~MoTe$_2$~~ & ~~WS$_2$~~ & ~~WSe$_2$~~ & ~~WTe$_2$~~ \\
        \hline
        $\kappa_{xx}^{P}$(300~K)~~ & 12.16 & 11.26 & 12.40 & 10.22 & 10.95 & 12.00 \\
        $\kappa_{yy}^{P}$(300~K)~~ & 13.37 & 7.92  & 11.03 & 12.14 & 7.80 & 8.27 \\
        $\kappa_{xx}^{coh}$(300~K)~~ & 0.08 & 0.08 & 0.06 & 0.09 & 0.06 & 0.05 \\
        $\kappa_{yy}^{coh}$(300~K)~~ & 0.12 & 0.12 & 0.08 & 0.11 & 0.06 & 0.04 \\
        \hline\hline
    \end{tabular}
    \centering
    \renewcommand{\arraystretch}{1} 
    \caption{Particle- (Peierls) and wave-like (Wigner) contribution to the lattice thermal conductivity at room temperature within the Wigner-RTA. The Peierls component is computed within the RTA of the BTE.}
    \label{tab:wigner}
\end{table}

\subsection{Role of ultralow-frequency phonons in four-phonon collisions}
\label{sub:cutoff}

All the results presented above have been obtained with the \texttt{FourPhonon} package~\cite{HanCPC22}, an extension of ShengBTE~\cite{LiCPC14} that incorporates four-phonon scattering processes into the iterative solution of the phonon BTE beyond the RTA. \texttt{FourPhonon} excludes four-phonon scattering processes whenever the second, third, or fourth phonon involved in the collision has a frequency below 1.25 rad/ps ($\approx 0.2$~THz). This cutoff
prevents numerical artifacts arising from the slightly negative frequencies that can occur as the acoustic branches approach the $\Gamma$ point. In the present work, however, by enforcing the rotational invariance, we made sure than even the ZA mode, whose parabolic approach to $\Gamma$ is sometimes problematic, is well behaved. Consequently, this low-frequency cutoff is no longer required for numerical stability and can be safely reduced or even removed.

\begin{figure}[h]
\includegraphics[width=0.75\linewidth]{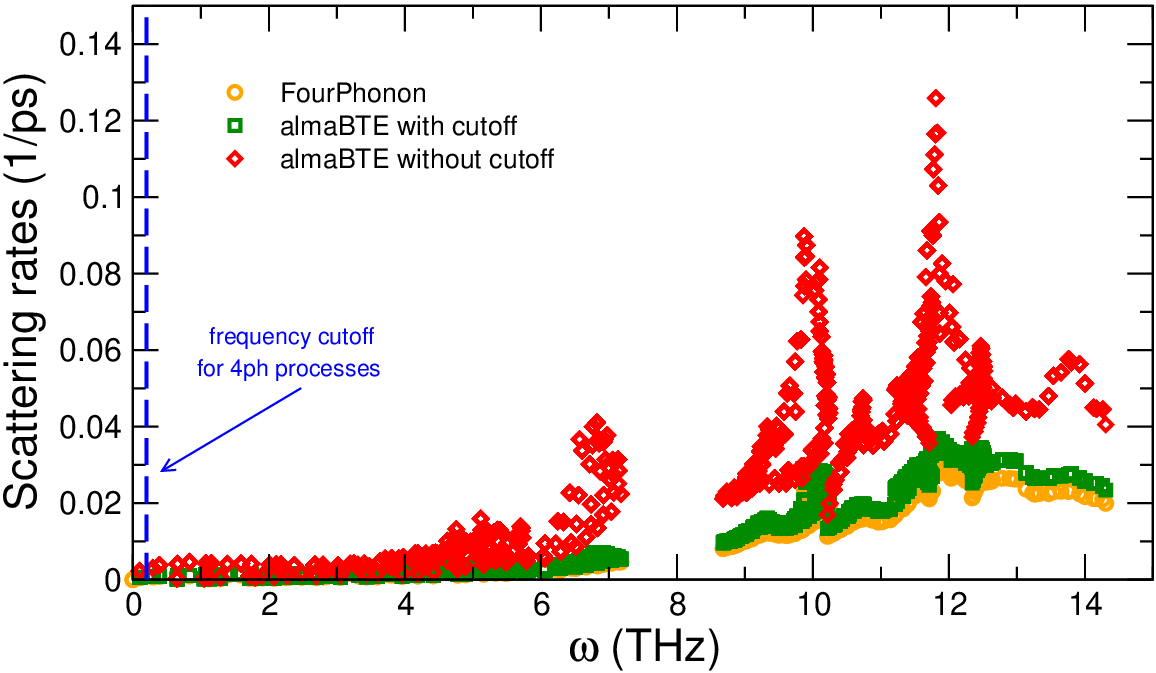}
\caption{4ph scattering rates for 2H-MoS$_2$ at 300 K computed with \texttt{FourPhonon} and \texttt{almaBTE}.}
\label{fig:cutoff}
\end{figure}

Indeed, we found that the inclusion of such ultralow-frequency phonons substantially enlarges the available 4ph phase-space, thus leading to a significant increase in the scattering rates. This can be seen in Figure~\ref{fig:cutoff}), where we compared the room-temperature 4ph scattering rates in 2H-MoS$_2$ obtained from (i)~from FourPhonon, with the built-in 1.25~rad/ps cutoff; (ii) the same in-house version of almaBTE, where we implemented the same cutoff, and (iii) almaBTE without any cutoff. The good agreement between (i) and (ii) indicates that the two codes yield very similar results when the same set of phonons are considered (i.e., those with $\omega > 1.25$~rad/ps). On the other hand, when this spurious cutoff is removed and phonons down to $\omega \rightarrow 0$ can correctly participate in 4ph collisions, the scattering rates increases, leading to a $\approx10$\% decrease in ${\bf \kappa}$ within the RTA (almaBTE RTA-results with and without cutoff are 102.00 and 91.42~W~m$^{-1}$K$^{-1}$, respectively).

Although they call for a more systematic investigation (over more materials, polymorphs, and temperatures) the results for 2H-MoS$_2$ of Figure~\ref{fig:cutoff} suggest that the reduction of ${\bf \kappa}$ ascribable to 4ph processes --which we found to be already sizeable in our analysis in Figure~\ref{fig:4phred} and \ref{fig:4phred_1Tp}-- can even be larger and is then important to account for it.


\subsection{Conclusions}
\label{sub:conc}

In summary, we have presented an exhaustive study of the lattice thermal conductivity of single-layer group-VIB TMDs, MX$_2$ (M= Mo, W, X = S, Se, Te), within an entirely {\it ab initio} theoretical framework, where we solve the Boltzmann Transport Equation (BTE) beyond the Relaxation Transport Approximation (RTA). In particular, we focused on the role of fourth-order phonon-phonon processes, often neglected. We showed that for a quantitative estimate of ${\bm \kappa}$ these processes must be accounted. Indeed, although their role is stronger at high temperatures, as one expects, they are responsible for an additional suppression of the lattice thermal conductivity ranging from 15\% to 35\% even at room temperature. Our study includes both the 2H --the ground-state of all six materials considered, but WTe$_2$-- and the \1Tp\ polymorph --thermodynamically competitive with the 2H, especially for selenides and tellurides. In all six single-layet TMDs the more symmetric 2H phase is more conductive than the \1Tp\ phase. Also, the importance of 4ph scattering in \1Tp\ polymorphs is considerably lower, as third-order phonon-phonon collisions are much more dominant than in the 2H phase.

The possibility of dynamical phase engineering, where 2H$\leftrightarrow$\1Tp\ phase transitions are triggered by external fields paves the way for the design of phonon-based logic elements. Our results indicate that $\kappa_{1T^\prime}/\kappa_{2H}$ switching ratios of the order of 0.5 at room temperature can be achieved for tellurides, where field-controlled phase transitions have already been demonstrated experimentally. Sulphides and selenides are characterized by even lower $\kappa_{1T^\prime}/\kappa_{2H}$ ratios. TO date, however, the energy barrier that must be overcome for the 2H$\rightarrow$\1Tp\ transformation has proven to be sufficiently high to prevent the dynamical switching of one polytype into the other.

Finally, we have verified that the coherence component of the lattice thermal conductivity in the Wigner transport regime is negligible (i.e., several orders of magnitude smaller) compared to the usual particle-like treatment within the Peierls BTE, as expected in well-ordered  crystals.

\begin{acknowledgement}
We acknowledge financial support by MCIN/AEI/10.13039/501100011033 under grant PID2024-162811NB-I00, and the Severo Ochoa Centres of Excellence Program under grant CEX2023-001263-S. This work is also supported by the Horizon Europe research and innovation program of the European Union under the Marie Skłodowska–Curie grant agreement 10181337 (DOCFAM+). The work of ML has been carried out within the PhD program in Physics of the Universitat Aut\`onoma de Barcelona (UAB). Qi Ren is supported by the China Scholarship Council (CSC) (Grant No. 202506030153), the Postdoctoral Fellowship Program (Grade B) of China Postdoctoral Science Foundation (Grant No. GZB20250965) and the fellowship from the China Postdoctoral Science Foundation (Grant No. 2026M794611). Calculations were performed at the Centro de Supercomputaci\'on de Galicia (CESGA) and at the Barcelona Supercomputing Center (BSC) within actions FI-2025-2-0076, QHS-2025-2-0030, QHS-2025-3-0040, QHS-2026-1-0047 of the Red Espa\~nola de Supercomputaci\'on (RES). M.R.-M. acknowledges the valuable discussions with Jes\'us Carrete regarding the implementation of new features in \texttt{almaBTE}. 
\end{acknowledgement}




\providecommand{\latin}[1]{#1}
\makeatletter
\providecommand{\doi}
  {\begingroup\let\do\@makeother\dospecials
  \catcode`\{=1 \catcode`\}=2 \doi@aux}
\providecommand{\doi@aux}[1]{\endgroup\texttt{#1}}
\makeatother
\providecommand*\mcitethebibliography{\thebibliography}
\csname @ifundefined\endcsname{endmcitethebibliography}
  {\let\endmcitethebibliography\endthebibliography}{}

\end{document}